\documentstyle [12pt]{article}
\begin{document}

\begin{center}
P.S.Isaev\\[0.5cm]
{\bf "On New Physical Reality"\\
 (on $\Psi$--ether)}
\end{center}
\vspace*{1cm}

It is shown that there exists a new physical reality --
the $\Psi$--ether.
All the achievements of quantum mechanics and quantum field theory are due
to the fact that both the theories include the influence of
$\Psi$--ether on the physical processes occurring in the Universe.

Physics of the XXth century was first of all the physics of $\Psi$--ether.
\vspace*{0.5cm}

{\bf Introduction}
\vspace*{0.5cm}

\begin{tabular}{p{4cm}p{10cm}}
\it &
~~~ "\ldots When you follow any of our physics too far, you find that it
always gets in some kind of trouble\ldots " (R.Feynman, R.Leighton, M.Sanders
"Feynman lectures on physics", vol. II, p.28--1, Addison--Wesley Publishing
Company, INC. Reading, Massachusetts. Palo Alto. London, 1964)

"\dots There are so many things about elementary particles that we
still don't understand\dots " (ibid, p.28--5).\\
\end{tabular}
\vspace*{0.5cm}

As soon as the Einstein special theory of relativity was created, the
existence of the Maxwell--Lorentz ether was no longer discussed. The world
scientific community accepted the Einstein point of view: "There is no room
for the ether in the special relativity". At school lessons and university
courses of physics departments, children and students were taught during
the XXth century that the Maxwell--Lorentz ether does not exist, and
there exist only "fields" in vacuum. However, many physicists including the
world famous ones (Dirac, Schwinger) have not eliminated the possibility for
ether to exist.

Dirac in the paper "The evolution of the physicist's picture of nature"
[1] makes a series of critical remarks concerning the modern physics. He
considers that : 1) The requirement of four-dimensional symmetry of the world
is not obligatory. "A few decades ago, it seemed quite certain that one had
to express the whole of physics in four-dimensional form. But now it seems,
that four-dimensional symmetry is not of such overriding importance, since
the description of nature sometimes gets simplified when one departs from
it";
2) "We are led to a theory that does not predict with certainty what is going
to happen in the future but gives us information only about the probability
of occurrence of various events. This giving up of determinacy has been a very
controversial subject, and some people do not like it at all. Einstein in
particular never liked it. Although Einstein was one of the great
contributors to the development of quantum mechanics, he still was rather
hostile to the form that quantum mechanics evolved into during his lifetime
and that it still retains";
3) "Everyone is agreed on the formalism. It works so well that nobody can
afford to disagree with it. But still the picture that we are to set up
behind this formalism is a subject of controversy"
4) "The physics of the future, of course, cannot have the three quantities
$\hbar$, $e$ and $c$ all as fundamental quantities. Only two of them can
be fundamental, and the third must be derived from those two". In Dirac's
opinion, the quantities $e$ and $c$ will remain fundamental quantities, and
the quantity $\hbar$ will be expressed via $e$ and $c$;
5) "I am inclined to suspect that the renormalization theory is something
that will not survive in the future, and that the remarkable agreement
between its results and experiment should be looked on as a fluke".

Further Dirac writes: "I might perhaps discuss some ideas I have had about
how one can possibly attack some of these problems. None of these ideas has
been worked out very far, and I do not have much hope for anyone of them
\ldots ".

I) "\ldots One of these ideas is to introduce something corresponding to the
luminiferous ether which was so popular among the physicists of the 19th
century \ldots When I talk about reintroducing the ether, I do not mean to
go back to the picture of the ether that one had in the 19th century, but I
do mean to introduce a new picture of the ether that will conform to our
present ideas of quantum theory" \ldots

II) "The picture I propose goes back to the idea of Faraday lines
\ldots We can suppose that the continuous distribution of Faraday lines o
forces that we have in the classical picture is replaces by just a few
discrete lines of force with no lines of force between them \ldots
We can picture the lines of force as strings" \ldots

III)"I might mention a third picture with which I have been dealing lately.
It involves departing from the picture of the electron as a point and thinking
of it as a kind of sphere with a finite size"\ldots

Schwinger also tries to revive the idea of the ether. As an epigraph to the
paper "A Magnetic model of matter"[2], Schwinger takes a quotation from "The
Mathematical Principles of Natural Philosophe" by Newton: "And now we might
add something concerning a certain most subtle spirit, which pervades and
lies hid in all gross bodies". Schwinger assumes dual-charged particles as a
basis of the magnetic model of matter. And, perhaps, Schwinger considered it
possible to construct the Newton ether mentioned in the epigraph to his
paper from "dyons"---elementary constituents of magnetic matter. This attempt
by Schwinger demonstrates him being dissatisfied with the state of physical
theory at his time.

How was the ether conceived at the end of the XIXth century? A rather
complete idea of it was given by Maxwell. He said [3]: "The vast
interplanetary and interstellar regions will no longer be regarded as waste
places in the universe, which the Creator has not seen fit to fill with the
symbols of the manifold order of His kingdom. We shall find them to be
already full of this wonderful medium; so full that no human power can remove
it from the smallest portion of space, or produce the slightest flaw in its
infinite continuity.  It extends unbroken from star to star; and when a
molecule of hydrogen vibrates in the dog-star, the medium receives the
impulses of these vibrations; and after carrying them in its immense bosom
for three years , delivers them in due course, regular order, and full tale
into the spectroscope of Mr. Huggins, at Tulse Hill.

But the medium has other functions and operations besides bearing light from
man to man, and from world to world, and giving evidence of the absolute
unity of the metric system of the universe. Its minute parts may have
rotatory as well as vibratory motions, and the axes of rotation form those
lines of magnetic force which extend in unbroken continuity into regions
which no eye has seen, and which, by their action on our magnets, are telling
us in language not yet interpreted, what is going on in the hidden underworld
from minute to minute and from century to century".

In the paper "Ether" [4], Maxwell wrote: "\dots Ether is a material substance
of a more subtle kind than visible bodies, supposed to exist in those parts
of space which are apparently empty \dots Newton himself, however,
endeavoured to account for gravitation by differences of pressure in an ether
(see Art. Attraction*, Vol. III, p. 64); but he did not publish his theory,
"because he was not able from experiment and observation to give a
satisfactory account of this medium, and the manner of its operation in
producing the chief phenomena of nature." \dots We therefore conclude that
light is not a substance but a process going on in a substance, the process
going on in the first portion of light being always the exact opposite of the
process going on in the other at the same instant, so that when the two
portions are combined no process goes on at all".

In the Maxwell opinion, the ether possesses elasticity, hardness, finite
density, but it is different from the usual matter. The vibrations of light
are transverse. "\dots We know that the ether transmits transverse vibrations
to very great distances without sensible loss of energy by dissipation \dots
If there is any motion of rotation, it must be a rotation of very small
portions of the medium each about its own axis, so that the medium must be
broken up into a number of molecular vortices \dots" [5].

So, according to Maxwell, "the unity of the metric system of the universe" is
inseparable from the idea of the ether he has introduced: the ether is
functioned as a carrier of the electromagnetic field, and possibly, of the
gravitational field, and light is not a substance but a process going on in a
substance\dots and if there is any motion of rotationin the ether it must be a
rotation of very small portions of the medium each about its own axis, so
that the medium must be broken up into a number of molecular vortices.

However, the Michelson experiments (1881) and the Michelson--Morli
experiments (1885--1887) [M; M--M] have shown that there is no "ether wind"
when the Earth moves through the ether, and it turned out to be impossible to
reveal the motion of the Earth through the ether with an accuracy up to terms
$\frac{V^2}{c^2}$ where $V$ is the velocity of the Earth, and $c$ is the
velocity of light. A poetic picture of the ether presented by Maxwell was
either a failure or required a serious theoretical substantiation.
In order to make the hypothesis of stationary ether be consistent with the
negative result of the experiments [M; M--M] Fitzgerald and Lorentz put
forward the hypothesis of contraction of bodies in the direction of their
motion by a factor $\sqrt{1-\frac{V^2}{c^2}}$.

In 1895 [6], Lorentz said that this hypothesis is not so strange may if we
assume that molecular forces are also transferred through the ether.
The form and sizes of a solid body are ultimately conditioned by the strength of molecular
interactions, so the sizes be changed in this case; therefore, from a
theoretical point of view, there are no objections against this hypothesis.

In 1905, Poincare wrote that Lorentz considered it necessary to supplement
his hypothesis so that the postulate of relativity hold valid also in the
presence of other forces in addition to magnetic ones. According to Lorentz,
all the forces irrespective of their origin, owing to the Lorentz
transformation (and, consequently, owing to translational motion),
behave just in the same way as electromagnetic forces do. It turned out
necessary to examine this hypothesis more carefully and to investigate which
changes are introduced by it in the laws of gravity.

If propagation of the gravitational forces occurs with the velocity of light,
this cannot be a result of any random circumstances, this should rather be
conditioned by one of the functions of the ether, and then, there arises the
problem of a deeper investigation of the nature of that function and to
connect it with other properties of the either [7].

Poincare also wrote that the  Fitzgerald--Lorentz hypothesis seem strange
at first sight. All we can say at present in favor of it is that it is an
immediate expression of the Michelson experimental results if we determine
distances by the time necessary for light for them being covered.

It is impossible to get out of impression that the relativity principle is a
universal law of the nature. We can never discover anything except for
relative velocities. By this I mean not only the velocities of bodies with
respect to the ether, but also the velocities of bodies with respect to each
other [8].

In 1912, in the paper "Hypothesis of quanta", Poincare wrote that none of
experiments can reveal whether a body is at rest or in absolute motion
with respect to the absolute space or with respect to the ether [9].

So, Newton, Faraday, Fizeau, Maxwell, Lorentz, Poincare, Plank, and others (I
cannot cite all the names, of course) considered the ether to be a real
substance in their studies.

Poincare proved the group properties of Lorentz transformations; Minkowski
introduced the 4-dimensional space--time. In this way, there arose the
relativistic representation of basic laws, Lagrangians, and equations of
motion (and even the relativistic form of calculations) in modern theory.
Relativity arose and was substantiated on the basis of existence of the ether.

The Einstein special theory of relativity disclaims the existence of the
ether. In 1905, Einstein in the  paper " On the Electrodynamics of Moving
Bodies" wrote [10]:
"\ldots the unsuccessful attempts to discover any motion of the earth relatively
to the "light medium," suggest that the phenomena of electrodynamics as well
as of mechanics possess no properties corresponding to the idea of absolute
rest. They suggest rather that, as has already been shown to the first order
of small quantities, the same laws of electrodynamics and optics will be
valid for all frames of reference for which the equations of mechanics hold
good. (The precading memoir by Lorentz was not at this time known to the
author.) We will raise this conjecture (the purport of which will hereafter
be called the "Principle of Relativity") to the status of a postulate, and
also introduce another postulate, which is only apparently irreconcilable
with the former, namely, that light is always propagated in empty space with
a difinite velocity $c$ which is independent of the state of motion of the
emitting body. These two postulates suffice for the attainment of a simple
and consistent theory of the electrodynamics of moving bodies based on
Maxwell's theory for stationary bodies. The introduction of a "luminiferous
ether" will prove to be superfluous inasmuch as the view here to be developed
will not require an "absolutely stationary space" provided with special
properties\ldots"

It is appropriate to accompany this statement with the question: How one can
construct the dynamics of moving bodies on the basis of the Maxwell theory for
bodies at rest if in special theory of relativity there is no definition of
a body at rest?

At the same time, Einstein said that the absolute space in the Newton
mechanics is just the ether and that only by introducing the conception of
ether as absolute space, one can distinguish the uniform linear motion from
the accelerated, rotational motion.

Then, one more question arises: The relativistic mechanics in the limit
of small velocities reduces to the Newton mechanics. Which is the origin of
the ether recognized by Einstein in the Newton theory if special theory of
relativity disclaims its existence?

Einstein said the influence of the ether in the Newton theory is included
into the formulation of laws of the Newton dynamics,and therefore, the space
according to Newton turned out to be absolute and empty. Analogous situation
arose with including the ether into the second postulate of special
relativity about constancy of the "velocity $c$ which is independent of the
state of motion of a emitting body", and the ether turned out to be unnecessary , and
the space became empty.

It is to be stressed that Einstein was not an opponent of the ether. In his
speech "Ether and relativity theory" at the Leiden university on May 5, 1920,
on the occasion of his election as honorary Professor, he said that the
recognition of the fact that the empty space is not uniform and isotropic in
physical aspects forces us to describe its state with ten functions---
gravitational potentials $q_{\mu \nu}$. However in doing so, the notion of
the ether again acquires a definite content  that is completely different
from the content of the concept of the mechanical theory of light. The
ether of general relativity is a medium that is deprived of all mechanical
and kinematic properties, but at the same time, defines mechanical (and
electromagnetic) processes. Also, he said that we know that it defines metric
relations in the space--time continuum (see ref. [11]).

So, we are faced with two points of view:

1) Lorentz and Poincare think that the Lorentz transformations are an
immediate expression of the Michelson experimental results if distances are
defined by the time necessary for light to cover them. So, the Lorentz
transformations reflect the properties of a medium in which light propagates
and include the influence of the medium (ether) on physical properties
occurring in the Universe.

2) In special theory of relativity it is considered that there is no ether,
but there is an empty space, and that light in vacuum always propagates with
a definite velocity $c$ independent of the state of motion of a emitting
body. As we see, this postulate of special theory of relativity expresses
in words the physical content of the Lorentz transformations published
earlier.

From comparison of these points of view it follows that the relativistic form
of writing all the laws and equations of elementary particle physics can be
considered, on the one hand, as a way of including the influence of the ether
on physical processes occurring in microcosm, but, on the other hand, since
the influence of the ether is taken into account in the process of
relativization of formulae, then it is as though the ether is absent, it does
not exist , as special theory of relativity states.

In this paper, I advocate the thesis that the ether as a physical reality
does exist, show how the influence of the ether entered into formulae of
quantum mechanics and quantum field theory, and construct a model of the
real $\Psi$--ether.

The model of the  $\Psi$--ether I proposed here satisfies modern requirements
of quantum field theory, as was said by Dirac. \underline {The $\Psi$--ether
is defined as the Bose--Einstein condensate}\\ \underline { of neutrino--antineutrino pairs
of the Copper type.}

How can one arrive at this definition? The Klein--Gordon relativistic
equations for a scalar particle with mass $m$ without interaction (or, as it
is written in books at present, in vacuum - emptiness) are of the form:
$$
\Bigl( \hbar^2\nabla^2 - \frac{\hbar^2}{c^2}\frac{\partial^2}{\partial t^2} - m^2c^2\Bigr)\Psi (x,t) = 0\, ,
$$
\noindent
where
$$\nabla^2 = \frac{\partial^2}{\partial x^2}+
\frac{\partial^2}{\partial y^2}+\frac{\partial^2}{\partial z^2};
\ \Box \equiv \frac{\partial^2}{\partial x^2}+
\frac{\partial^2}{\partial y^2}+\frac{\partial^2}{\partial z^2} -
\frac{\partial^2}{c^2\partial t^2}
$$
\vspace*{0.5cm}

The equation for a vector field for a particle with mass $m$ is written in
the same form, but $\Psi (x,t)$ is now a multicomponent function.

Since I assert that any relativistic equation for a free particle with mass
$m$ should be understood not as an equation in vacuum but as an equation for
a particle with mass $m$ in the ether, then setting the mass $m$ equal to
zero, we obtain the simplest equation for the ether described by the scalar
function $\Psi (x,t)$:
\begin{equation}
\Biggl[ \hbar^2\Biggl( \frac{\partial^2}{\partial x^2} + \frac{\partial^2}{\partial y^2} + \frac{\partial^2}{\partial z^2}\Biggr)
- \frac{\hbar^2}{c^2} \frac{\partial^2}{\partial t^2}\Biggr] \Psi (x,t) = 0.
\end{equation}
\vspace*{0.3cm}

Hence, it is clear why I call my model of the ether the model of
$\Psi$ --ether: first, in a language conventional for physicists, the ether
is described by the $\Psi (x,t)$ -function; second, as we will see below, the
$\Psi$--ether
differs in its properties from the Lorentz--Maxwell ether, and it is
necessary to supply it with a special name.

Consider the case of the electromagnetic field. We denote the strengths of
electric and magnetic fields, respectively, by $\vec{E}$ and $\vec{H}$[12].
If we introduce the vector $\vec{A}$ and scalar $\varphi$ potentials with the use of the
relations:

\begin{eqnarray}
\vec{H} & = & rot \vec{A}  \nonumber\\
\vec{E} & = & -\frac{1}{c} \frac{\partial \vec{A}}{\partial t} - qrad \varphi ,
\end{eqnarray}

\noindent and make use of the Lorentz condition
\begin{equation}
div \vec{A} + \frac{1}{c} \frac{\partial \varphi}{\partial t} = 0,
\end{equation}

\noindent then the potentials $\vec{A}$ and $\vec{\varphi}$,
as it is known, obey the equations
\begin{eqnarray}
\Box \vec{A} & = & \nabla^2\vec{A} - \frac{1}{c^2} \frac{\partial^2\vec{A}}{\partial t^2} = 0 \nonumber \\
\Box \varphi & = & \nabla^2\varphi - \frac{1}{c^2} \frac{\partial^2 \varphi}{\partial t^2} = 0\, .
\end{eqnarray}

\noindent In limits of the Lorentz gauge (3), the scalar potential can be taken zero.
Then the charge-independent part of the potentials $\vec{A}$ and $\varphi$
is determined by the equation:

\begin{eqnarray}
\Box \vec{A} & = & 0; \quad div \vec{A} = 0; \quad \varphi = 0; \nonumber \\
\vec{E} & = & -\frac{1}{c} \frac{\partial \vec{A}}{\partial t}; \quad \vec{H} = rot \vec{A}.
\end{eqnarray}

The system of equations (5) turns out to be completely
equivalent to the Maxwell--Lorentz equations. In this case, the general
solution of equations (5) is given by a superposition of transverse waves [12].

The vector $\vec{A}$ and scalar $\varphi$ potentials for a more symmetric representation
can be combined into one four-dimensional vector $\Psi_i$.
Then equations (4) are written in the form:
\begin{equation}
\Box \Psi_i(x,t) = 0.
\end{equation}

I define these equations as equations for the real $\Psi$--ether.

I would like to emphasize something unusual in the transition from the
Maxwell equations to the equations for electromagnetic potentials
$\vec{A}$ and $\varphi$,
i.e., to equation (6) for the $\Psi$--ether.

No attention is paid in any textbook or monograph to the fact that
introducing auxiliary electromagnetic potentials $\vec{A}$ and $\varphi$
by formulae (2), we equate physical observables $\vec{H}$ and $\vec{E}$
to nonphysical auxiliary functions $\vec{A}$ and $\varphi$.
This is impermissible in physics! If one hand-side of an
equation contains a physical observable, the other should also contain a
physical observable.

As a matter of fact, the potentials $\vec{A}$ and $\varphi$ in relation (2)
describe a physical reality---it is the $\Psi$--ether, and this physical
reality is observable.

The Klein--Gordon operator can be represented as a product of two commuting
matrix operators
$$
I_{\alpha \beta}(\Box - m^2) = \sum_{\delta}\Bigl( i\gamma^n \frac{\partial}{\partial x_n}+m\Bigr)_{\alpha \delta}\Bigl(i\gamma^k \frac{\partial}{\partial x_k}-m\Bigr)_{\delta \beta}
$$
and in order that the field function obey the Klein--Gordon equation
$$
(\Box - m^2)\Psi = 0,
\eqno{(a)}
$$
we can require that it satisfy also one of the first-order equations
$$
\Biggl( i\gamma^n \frac{\partial}{\partial x_n} +m\Biggr)\Psi = 0 \qquad {\rm or} \qquad
\Biggl( i\gamma^n \frac{\partial}{\partial x_n} - m\Biggr)\Psi = 0.
\eqno{(b)}
$$
Equations (b) are less general than equation (a), and though any solution of
one of equations (b) satisfies equation (a), the reverse does not hold.

A common property of all solutions of the Klein--Gordon equation
corresponding to single-valued representations of the Lorentz group is that
the corresponding particles possess integer spins (0 1, ...). Particles with
a half-integer spin are described by a spinor representation [13]. Putting
$m=0$ in equations (b), we derive equations for the neutrino--antineutrino
field. So, returning to relations (2) and (6), we can assert that the
electromagnetic potentials are not merely auxiliary functions, but they are
connected by definite relations with the physical reality - $\Psi$--ether--by
neutrino-antineutrino pairs.

The $\Psi$--ether forms a superconducting medium filling the whole world space,
does not manifest itself in the heat capacity of bodies, permits
propagation of transverse waves (it is a carrier of electromagnetic waves)
and of spinor and boson waves (it is their carrier), does not prevent the
motion of elementary particles,nuclei, and cosmic bodies including stars in
the world space. We are fairly aware of that the neutrino (a neutrino wave)
penetrates the thicknesses of stars, the Sun, and the Earth without
essential change of the direction and energy losses. It can be assumed that
the scalar component of the $\Psi$--ether serves as a transmitter of
gravitational forces. We can add to all said above that a continuous
oscillating system is dynamically equivalent to a set of harmonic
oscillators. There is the Weyl--Kurant theorem: If the wavelength is small
as compared to linear sizes of an oscillating system, and limiting conditions
are uniform, the frequency spectrum does not depend on a particular form of
limiting conditions and on the shape of a cavity.

The model of $\Psi$--ether as a continuous oscillating system obey the conditions
of the Weyl--Kurant theorem, and the $\Psi$--ether can, in principle, be
considered as a set of harmonic oscillators.

Now, let us point to an indissoluble connection of the $\Psi$--ether with
quantum mechanics.

Let us recall how the problem of a hydrogen atom is solved in quantum
mechanics. When the potential energy $V(\vec{r})$ is time-independent and
spherically symmetric, so that $V(\vec{r})$ depends only on the absolute
value $r$ of vector $\vec{r}$ the solution to the schr\"oedinger wave equation
\begin{equation}
-i\hbar \frac{\partial \Psi}{\partial t} = - \frac{\hbar^2}{2m} \nabla^2 \Psi + V(\vec{r},t)\Psi
\end{equation}
\vspace*{0.2cm}

is derived by separating the variables
\begin{eqnarray}
\Psi (\vec{r},t) & = & u(\vec{r})f(t) \nonumber \\
u(\vec{r}) & = & u(r,\theta , \varphi ) = R(r)Y(\theta , \varphi ).
\end{eqnarray}
Since the problem of a hydrogen atom is considered as the problem of two-body
interaction (a proton and an electron) and is solved with the purpose to
find out stationary states with a constant value of energy $E$, so that
$\Psi (x,t) = Ce^{-i \frac{Et}{\hbar}}$, where $C$ is a normalization
constant, then for the functions $R(r)$ and $Y(\theta , \varphi )$
from equation (7) we derive the following equations [14]:
\begin{equation}
\frac{1}{\sin \theta} \frac{\partial}{\partial \theta}\Bigl( \sin \theta \frac{\partial Y}{\partial \theta}+
\frac{1}{\sin^2\theta} \frac{\partial^2Y}{\partial \varphi^2}\Bigl)+\lambda Y = 0
\end{equation}
\begin{equation}
\frac{1}{r^2} \frac{d}{dr}\Bigl(r^2\frac{dR}{dr}\Bigr)+\Big\{ \frac{2\mu}{\hbar^2}[E-V(r)]-\frac{\lambda}{r^2}\Big\}R = 0
\end{equation}
\vspace*{0.3cm}
where $\lambda$ is the constant of separation of variables in the functions
$R(r)$ and $Y(\theta ,\varphi )$, $\mu$ is the reduced mass of the system
of proton (p) + electron (e); $E$ is the energy of a level for the bound
state $p+e \ (E<0)$, $V(r)$ is the potential energy of proton--electron
interaction equal to $\frac{e^2}{r}$. Equation (9) is also solved by
separating the variables
$$
Y(\theta , \varphi ) = \Theta (\theta )\Phi (\varphi ),
$$
as a result of which, we arrive at the equations
\begin{equation}
\frac{\partial^2\Phi (\varphi )}{\partial \varphi^2}+\nu \Phi (\varphi ) = 0
\end{equation}
\begin{equation}
\frac{1}{\sin \theta} \frac{d}{d\theta}\Bigl(\sin \theta \frac{\partial \Theta (\theta )}{\partial \theta}\Bigr)+
\Bigl(\lambda - \frac{\nu}{\sin^2\theta}\Bigr)\Theta (\theta ) = 0
\end{equation}
\vspace*{0.3cm}

The solution for $\Phi (\varphi )$ is of the form
$$
\Phi_m(\varphi ) = \frac{1}{2\pi}e^{im\varphi}; \quad \nu = m^2
$$
(the value of $"m"$ can be any integer positive or negative number); and
physically admissible solutions to the equation for the function
$\Theta (\theta )$ are those that satisfy the conditions
$$
\lambda = l(l+1)
$$
$$
|m| \leq l.
$$
These solutions are expressed through $\Theta (\theta )$ the associated
Legendre polynomials. In the end, the radial part $R(r)$
of the Schr\"oedinger equation obeys the following equation:
\begin{equation}
\frac{1}{r^2} \frac{d}{dr}\Bigl(r^2 \frac{dR}{dr}\Big)+ \frac{2\mu}{\hbar^2} \frac{e^2}{r} R(r) +
\frac{2\mu}{\hbar^2}E R(r) - \frac{l(l+1)}{r^2}R = 0
\end{equation}
Here, I draw attention to one of the critical moments of quantum mechanics.

The radial equation (13) is one-dimensional in which the potential energy
$V(r)$ is dependent on two parts
\begin{equation}
V(r) = + \frac{2\mu}{\hbar^2} \frac{e^2}{r} - \frac{l(l+1)}{r^2} .
\end{equation}
If the term $\sim \frac{e^2}{r}$ is responsible for the Coulomb interaction
of a proton with an electron in a hydrogen atom, the second term
$\frac{l(l+1)}{r^2}$ does not depend on any physical interaction.
The latter term originates from angular variables of
the wave function. Nevertheless, Schiff wrote: "The additional "potential
energy" can be seen physically to be connected with the angular momentum" [14].

However, if one puts the Coulomb interaction $\frac{e^2}{r}$
in the radial equation (13) to be zero, so that there is no interaction
between a proton and an electron, the term $\frac{l(l+1)}{r^2}$
does not disappear, and it makes no sense to relate its origin
with the angular momentum.

In fact, the term $\frac{l(l+1)}{r^2}$ appears in equation (13) because
of its $\Psi$--ether origin. To verify this, let us turn to the theory of
wave guides.

Let us consider the problem of finding the natural electromagnetic vibrations
of a hollow sphere resonator [15].

For the Borgnis function $U(r,\theta ,\varphi )$, we can write the equation
\begin{equation}
\frac{\partial^2U}{\partial r^2} + \frac{1}{r^2\sin \theta}\Biggl[ \frac{\partial}{\partial \theta}
\sin \theta \frac{\partial U}{\partial \theta} + \frac{\partial}{\partial \varphi} \frac{1}{\sin \varphi}
\frac{\partial U}{\partial \varphi}\Biggr] +k^2U = 0.
\end{equation}
The Borgnis function is connected by definite relations with the electric
$\vec{E}$ and magnetic $\vec{M}$ fields, and when it obeys equation (15), the
Maxwell equations hold also valid. On the other hand, the function $U$ is
connected by definite relations with the potentials $\vec{A}$ and $\varphi$,
i.e. with the $\Psi$--ether. Solution to equation (15) is derived like in
quantum mechanics by the method of separation of variables:
$$
U(r,\theta ,\varphi ) = F_1(r)F_2(\theta ,\varphi )
$$
\vspace*{0.2cm}

(I keep the notation from the monograph by de Broglie deliberately.)
The functions $F_1(r)$ and $F_2(\theta ,\varphi )$ satisfy the equations
$$
\frac{1}{\sin \theta} \frac{\partial}{\partial \theta} \sin \theta \frac{\partial F_2}{\partial \theta} +
\frac{1}{\sin^2\theta} \frac{\partial^2F_2}{\partial \varphi^2} + \gamma F_2 = 0
\eqno{(16a)}
$$
$$
r^2 \frac{\partial^2F_1}{\partial r^2} + k^2r^2F_1 - \gamma F_1 = 0
\eqno{(16b)}
$$
\vspace*{0.2cm}

In the problem, we consider electromagnetic waves harmonic in time and
characterized either by the frequency $\nu$
$$
\nu = \frac{kc}{2\pi}
$$
or by the wave vector
$$
k = \frac{2\pi \nu}{c}, \quad [k] = \frac{1}{cm}
$$
\vspace*{0.2cm}

Equation (16b) contains the quantity $k^2$, the quantity $\gamma$
in equation (16a)---the constant of separation of variables.

Equation (16a) completely coincides with equation (9), and solutions of eq.
(16a) are spherical functions. The regular solution of equation (16b) for
all $\Theta$ and $\varphi$ exists only when $\gamma = n(n+1)$.
Thus, equation (16b) assumes the form
\addtocounter{equation}{1}
\begin{equation}
\frac{d^2F_1}{dr^2} + \Bigl[k^2 - \frac{n(n+1)}{r^2}\Bigr]F_1 = 0
\end{equation}
\vspace*{0.3cm}

If we set $F_1(r)=rf(r)$, equation (17) is written in the form
\begin{equation}
\frac{d^2f}{dr^2} + \frac{2}{r} \frac{df}{dr} + \Bigl[k^2 - \frac{n(n+1)}{r^2}\Bigr]f(r) = 0.
\end{equation}
Upon simple computations, equation (13) from quantum mechanics can be written
as follows:
\begin{equation}
\frac{d^2R}{dr^2} + \frac{2}{r} \frac{dR}{dr} + \Biggl( \frac{2\mu E}{\hbar^2} +
\frac{2\mu e^2}{\hbar^2r} - \frac{l(l+1)}{r^2}\Biggr)R = 0.
\end{equation}
\vspace*{0.3cm}
If we put the term in (19) responsible for the Coulomb interaction of a
proton with an electron $\Biggl(= \frac{2\mu e^2}{\hbar^2r}\Biggr)$
equal to zero and replace $E$ by $E = \frac{p^2}{2\mu}$, equation (19) is
rewritten in the form
\begin{equation}
\frac{d^2R}{dr^2} + \frac{2}{r} \frac{dR}{dr} + \Biggl( k^2 - \frac{l(l+1)}{r^2}\Biggr) R = 0
\end{equation}
$$
\Biggl( \frac{2\mu p^2}{2\mu \hbar^2} = \frac{k^2\hbar^2}{\hbar^2} = k^2, \ k - {\rm wave \ vektor}\Biggr).
$$
\vspace*{0.3cm}

Equations (18) and (20) are identical and are solved under the same boundary
conditions: like in quantum mechanics, a solution to eq. (18) is sought
for $f(r)$ such that $f(r)$ be a finite function as $r\rightarrow 0$,
and when $r\rightarrow \infty$ the function $f(r)\rightarrow 0$
(on the boundary of a sphere). The corresponding solutions to equation
(18) describe standing waves inside the sphere at values
$$
n = 0,1,2..., \qquad m\leq n.
$$

Since electromagnetic waves are nothing else than oscillations of the
$\Psi$--ether, the term $\frac{n(n+1)}{r^2}$ in equation (18) is responsible
for standing waves of the $\Psi$--ether in a sphere resonator.

Thus, we conclude that the problem of finding the energy levels in a hydrogen
atom with the use of the Schr\"oedinger equation is in physical content
equivalent to the problem of finding natural electromagnetic oscillations
inside a sphere resonator; one of the basic postulates of quantum
mechanics---quantization of orbits in a hydrogen atom (the Bohr postulate
$mvr = \frac{nh}{2\pi}$)---is equivalent to the determination of conditions
for existence of standing waves of the $\Psi$--ether in a sphere resonator.

We see that quantum mechanics is equivalent to "mechanics" of the
$\Psi$--ether. The equation for the $\Psi$--ether is directly connected with
the Maxwell equations. The relativistic form of the equation for $\Psi$--ether
is in all the equations of elementary particle physics and in all the
Lagrangians of quantum field theory. So, the physics of XXth century has been
physics of the $\Psi$--ether. The genius of Maxwell anticipated almost all
properties of the ether but that the ether consists of neutrino--antineutrino
pairs ($\nu \tilde{\nu}$-- pairs). However, we know that the neutrino was
discovered experimentally only in 1953--1956 by Raines and Coen. Especially
surprising is the Maxwell genius prevision in the following aspects:

1) Smallest parts of this medium can have not only oscillatory motions
but also rotational ones, the axes of rotation being the corresponding
magnetic lines of force. If there exists the rotational motion, it should be
the rotation of very small parts of the medium, each around its own axis, so
that the medium should disintegrate into a great number of molecular
vortices.
At present, we know that the neutrino has spin $\frac{\hbar}{2}$, so that the Maxwell
prediction is justified.

2) "\ldots We therefore conclude that light is not a substance but a process
going on in a substance, the process going on in the first portion of light
being always the exact opposite of the process going on in the other at the
same instant, so that when the two portions are combined no process goes on
at all". Indeed, the decay of a "$\Psi$--ether molecule" consisting of a
Cooper $\nu \bar{\nu}$-pair in the simplest case completely corresponds to
the Maxwell description in which, instead of light, one should mean physical
properties of the $\nu \bar{\nu}$-pair.

Schwinger in paper [2] called the fundamental dual-charged particle
introduced by him the "dyon". By analogy, I propose to call the
neutrino--antineutrino pair the "psyon" and use it in what follows.

3) The $\Psi$--ether being a homogeneous medium ensures the unity of measure
and number in the Universe, which cannot be done by various fields of quantum
field theory.

Now, we need not develop the picture of the world constructed in the
framework of the Standard Model. We construct another picture of the world.
The whole visible known Universe is immersed in the all-embracing and
all-penetrating $\Psi$--ether and lives, and develops according to its laws.
The $\Psi$--ether is an abyss in which the known physical world negligible as compared to the
ether is immersed. Which are the relations between these two worlds---our
physical world and the $\Psi$-ether---and which is the interaction between
them are still open questions to be solved in the new century or new
millenium.

The world consisting of protons, electrons and neutrinos is stable, we see
it and study it. In particular, we study the neutrino with the help of the
neutron and unstable particles. We do not see the world composed only of
neutrinos, but it can be supposed that it is infinite and diverse.

Until now, only a small part of the $\Psi$--ether properties is observed; in
particular, they can be seen in the phenomena of superfluidity and
superconductivity. The $\Psi$--ether takes part in formation of all the
chemical elements in the Universe, in production of all unstable elementary
particles, in the formation and possibly spontaneous appearance of life on
the Earth, brings back the energy, momentum, angular momentum that it obtains
from our physical world---we do not observe violation of the conservation laws of
these quantities. If the state of the ether surrounding us changes, then our
physical world immersed in it should also change so that to remain in a
certain "equilibrium" state with the ether. If parts, maybe giant, of the
$\Psi$--ether possess various energy states, and our Solar system can pass
from one layer of the $\Psi$--ether to another, then in the vicinity of the
Solar system including the Earth, there can occur dramatic phenomena of
global rise or fall of temperature of the climate, which can result in the
change of forms of life on the Earth, in heating or cooling the Earth
itself, etc.

Restoration of the $\Psi$--ether to its rights after its "expulsion" from
physics in special theory of relativity allows us to answer some critical
question posed earlier by Einstein and de Broglie:

1. The dualism "wave--particle" vanishes. The function $\Psi$ that describes,
in modern language, a "free" particle is actually a wave function of the
particle with take into account its motion through the ether; wave processes in the ether
accompanying a moving particle determine "wave properties" of the particle,
and when the particle collides, for instance with a thin metallic film, they
generate a diffraction pattern. It is not surprising that the probability of
finding a particle (an electron, a proton, etc.) after a collision with a
metallic film is defined by the square of the ether wave function
$\Psi$--ether (i.e. $|\Psi (x,t)|^2$), by analogy with light whose intensity
is given by the amplitude squared.

2. The very essence of all world processes is now defined by the $\Psi$--ether
interaction. All electromagnetic processes are proportional to an integer
value of the Planck constant $\hbar$, since an electromagnetic wave in the
$\Psi$--ether theory is described by the vector component of the $\Psi$--ether.

3. The plane--wave solution
$$
\Psi \sim e^{\frac{i}{\hbar}(Et-\vec{k}\vec{x})}
$$

should not be interpreted within the Born statistical approach, as it was
said in p.1. The particle remains localized in the wave, as de Broglie
supposed, and thus, the principle of determinism is restored on what insist
Einstein. Theoretical physics returns to the possibility of the Einstein
description of a single system rather than an artificial ensemble of single
systems. De Broglie wrote that prominent scientists Planck, Einstein, and
Schroedinger, who belong to founders and pioneers of quantum theory from the
moment of its creation, always rejected the purely probabilistic
interpretation that was subsequently acquired by quantum physics.

De Broglie said that he was disappointed with the reaction, to his theory, of
other physicists--theorists seduced by the purely probabilistic
interpretation of Born, Bohr, and Heisenberg, refused his attempt, and
in subsequent years adhered to the generally accepted interpretation [16].

4. One more prediction by Einstein was realized: The Nature does not require
our choice between quantum and wave theory, it only requires the synthesis of
these theories, which is not yet achieved by physicists. In the proposed
model of the $\Psi$--ether this synthesis is obtained.

Acknowledgment of the $\Psi$--ether allows a new look onto a series of physical
phenomena in microcosm and astrophysics, extends the understanding of
physical, chemical, and biological processes in the Nature, and points to
some limits of our cognition.

1. The relict radiation can now be interpreted as age--long luminescence of a
weakly excited world $\Psi$--ether, or in other words, as eternal oscillations
of psyons forming the ether.

2. Equations of the type $\Box \Psi_i = 0 \ (a)$ and $(\Box - \mu^2)\varphi_i = 0$ (b)
are today determining equations in elementary particle physics. It would be
desirable to elucidate how the mass of elementary particle arises. One of the
answers can be found in the monograph by de Broglie "Electromagnetic waves in
wave guides and hollow resonators". He said that all the considered waves are
characterized by the "propagation factor"
$$
P = e^{i(kct-k_zt)}, \quad {\rm where} \quad k^2 = k_z^2+\alpha^2
\eqno{(see\;formula\;2.75\;[17])}
$$

Every possible wave is characterized by one of the eigenvalues of constant
$\alpha$ corresponding to the type of considered wave guides. The photon wave
should correspond to $k=k_z$, i.e. to the propagation with velocity $c$. This
corresponds to the propagation of an electromagnetic wave in vacuum. But if
the electromagnetic wave is confined inside a wave guide, then $k$ and $k_z$
are connected by the relation (2.75) where $\alpha$ differs from zero and
equals one of its eigenvalues corresponding to the form of the wave guide
under consideration. From the point of view of wave mechanics, in this case,
all occurs as if the photon had a proper mass determined by the form of a
wave guide and a given eigenvalue $\alpha_i$. So, it can be said that in a
given wave guide, the photon can possess a number of possible proper masses.
[17].

Thus, if we assume that in the process of collision of protons and electrons
with other protons and electrons, in an excited ether, there arise "wave
guide" conditions of propagation of the $\Psi$--ether waves, then under certain
conditions, there appears the whole spectrum of masses of unstable particles.
The particles will consist only of protons, electrons, and neutrinos---
there are no other constituents in the Nature. It is not surprising (this is
an experimental fact!) that all the unstable particles and resonances
detected till now decay only into protons, electrons, and neutrinos (and
their antiparticles). It is natural that all the "wave guide" particles can
be classified according to the parameters of $\Psi$--ether wave guides. A
great number of elementary particles is no longer exotic and becomes a
trivial consequence of the properties of the $\Psi$--ether. This notion of
the origin of the mass spectrum of elementary particles and resonances
suggests that, first, their number can be arbitrarily large, much larger than
the number discovered until now, and, second, their search can hardly be of
the scientific interest that is attached to it at present in connection with
the ideology of the standard model (this naturally concerns the search of the
Higgs boson).

We have considered one of the possible models of origin of masses of
elementary particles. It seems that it is difficult to explain in this way
the process of production of pairs of particles $e^+e^-,\ p\bar{p}$
and all other pairs fermions that occurs in processes of multiparticle production,
though the nature of psyons consisting of $\nu \bar{\nu}$--pairs suggests
to relate their decay with the production of $e^+e^-, \ p\bar{p}$ pairs and
others.

Clearly, the origin of symmetries in the world of elementary particles (for
instance, SU(3)-symmetries) is also determined by the symmetry properties of
the $\Psi$--ether. Establishment of the SU(3)-symmetry was not simple. I
recall that in the fifties when the mass of the
$\Lambda^0$--particle was thoroughly measured, experimenters assured that
there exist two values rather than one of the $\Lambda^0$--particle  mass,
i.e. it was assumed that there exist two different
$\Lambda^0$--particles, which was inconsistent with the SU(3)-symmetry model.

People told about the so-called "Eastern--Western effect". However, theorists
who believed in "sanctity" of SU(3)-symmetry insisted upon that there should
be only one $\Lambda^0$--particle. Nevertheless in booklets "Review of
Particle Properties" (especially in earlier publications), one can clearly
see a double-humped curve for the measured mass of the $\Lambda^0$--particle.

The KARMEN collaboration published data [18] which were interpreted by
Gninenko and Krasnikov [19] as the presence of two different modes of the
$\mu^+$--meson decay: the conventional mode:
$$
\ \mu^+ \rightarrow e^+ + \tilde \nu_e + \nu_{\mu}
$$
and an additional rarely encountered mode:
$\mu^+ \rightarrow e^+ + X$,
where $X$ is a new boson with mass 103.9 $MeV/c^2$.

In the $\Psi$--ether model, these two decay modes can be explained in a
natural way. Representing spins of $\mu , \ e$ and $\nu$ by arrows,
we can show the following two schemes of the $\mu^+$--structure:
\begin{tabbing}
\hspace*{1.5cm} \= \hspace*{1cm} \= \hspace*{1cm} \= \hspace*{1cm} \= \hspace*{1cm} \=
\hspace*{1.5cm} \= \hspace*{1.5cm} \= \hspace*{1cm} \= \hspace*{1cm} \=
\hspace*{1cm}  \= \hspace*{1cm} \=  \kill
\> $\uparrow$ \> $\rightarrow$ \> $\uparrow$ \> + \> $\uparrow \downarrow$ \> or \> $\uparrow$ \>
$\rightarrow$ \> $\downarrow$ \> + \> $\uparrow \uparrow$ \\[4mm]
\> $\mu^+$ \> \> $e^+$ \> \> $\nu_{\mu}\tilde{\nu_e}$ \> \> $\mu^+$ \> \>
$e^+$ \> \> $\nu_{\mu}\tilde{\nu_e}$ \\[4mm]
\>  \>  \> (a) \> \> \> \> \> \> (b) \> \> \\
\end{tabbing}

\vspace*{0.5cm}

In the scheme (a), the helicities of $\nu_{\mu}$ and $\tilde{\nu_e}$
coincide, and the pair $\nu_{\mu}\tilde{\nu_e}$ forms a more stable system
with a possible more delayed decay $\tau_{\mu}\sim 3.6 \mu s$,
than in the scheme (b), in which $\nu_{\mu}$ and $\tilde{\nu_e}$
have opposite helicities, and the pair $\nu_{\mu}\tilde{\nu_e}$
decays more rapidly with the lifetime of a three-body decay of $\mu^+$--meson
$\tau_{\mu}\sim 2.6 \mu s$. This is only qualitative considerations, and they
can be used for a qualitative analysis of the schemes of production and decay
of other unstable particles.

From this example it is seen that from the standpoint of the $\Psi$--ether
model, modern ideas of the structure components of matter should look
differently than it is postulated in the standard model.

In the standard model, structure components of matter are families of
quarks and leptons (fermions):
$$
\quad {u\choose d} \quad {c\choose s} \quad {t\choose b} \quad {e\choose {\nu_e}} \quad {\mu \choose {\nu_{\mu}}} \quad {\tau \choose {\nu_{\tau}}}
$$
\vspace*{0.3cm}

Interaction carriers are bosons: gluons, $\gamma$--quanta, $W$-
and $Z^0$--bosons.

The "pressure" of physicists-theorists on the development of the modern
elementary particle physics turned out to be so large, and the admiration for
the statement: "To be correct, a theory should be slightly mad" appeared to
be so great that the physical interpretation of phenomena within the
framework of the standard model went far beyond the limits of natural
understanding of the observed phenomena. Elementary particle physics becomes
more and more a hostage of mathematics. It is sufficient to point to the fact
that quarks "nonobservable in essence" and rapidly decaying $\mu$--and $\tau$-
-leptons are, on equal status with stable electrons and neutrinos, elevated to
the class of structure components of matter. Is this really that " mad spot"
of the standard theory which we intend to raise to the rank of "theory" after
the discovery of the Higgs boson?

As one of the most important properties of the $\Psi$--ether, we point out the
limits of validity of relativization of all formulae and calculations in
modern physics following from the structure of $\Psi$--ether. The principle of
relativization is essentially connected with the wave properties of the
$\Psi$--ether. Where the wave properties of the $\Psi$--ether come to an end,
there end electromagnetic waves, and this puts an end to our comprehension
of secrets of the Universe with the help of optical instruments and
radiotelescopes. The length of electromagnetic waves in the $\Psi$--ether is
limited from the side of both short and long waves.

From the side of short wavelengths, this limitation sets in when the
"mean free path" of psyons becomes smaller than the distance between psyons.
So, the density of psyons in $1 cm^3$ becomes decisive in the determination of
this critical limit. This density also determines the reliability of
information accepted and transferred by the $\Psi$--ether. If two or more
impulses come to a psyon, the gained information will be distorted in further
transmission.

From the side of large wave lengths, the limitations of relativization set
in when large wavelengths of the $\Psi$--ether lose their wave configuration
becoming a chaotic motion of huge masses of the $\Psi$--ether ("noises").

So, where the wave properties of the $\Psi$--ether come to an end, there
ends our cognition of secrets of the Universe through the electromagnetic
interaction. Man becomes blind and deaf. Evidently, there exist phenomena
whose description requires studies of the microscopic properties of
constituents of the $\Psi$--ether, psyons.

It is not clear which are the velocities of propagation of signals in the
$\Psi$--ether. If psyons can be arranged as strings, lattices, pyramidal
formations, cubes and parallelepipeds and so on and in the form of long chains, the
velocities of propagation of perturbations in them can be both larger and
smaller than the velocity of light. It is known that a neutrino signal from
the Supernova 1987 A came to the Earth several hours earlier that the signal
of light. In "Physics today" (April 1999, p. 9 and July 1999, p.17--18), it
was informed that the velocity of the light impulse passing through the
Bose--Einstein condensate lowers down to 17 m/sec, which is millions of times
lower than the light velocity in the ether. This slowing down of the velocity
of a light impulse can be produced not only by possible optical effects but
also by effects of the change of the ether structure in the experiments.

The above-assumed forms of combinations of neutrinos and antineutrinos in
various geometrical forms do not go beyond the scope of 3-dimensional space.
We will assume this picture of the space surrounding us.

Any changes of the state of the $\Psi$-ether in the Universe reach the
Earth, and consequently, the man from the moment of his conception up to
his death. Such sentences as "a man was  born under the Jupiter star",
or "yesterday I had a dream, but today...", or "I foresaw this..." can now be
substantiated not only psychologically, but also physically. It becomes
obvious that in the Nature, there can exist physical objects which cannot be
detected with the help of our five organs of sense, no matter how we
strengthen them. The man is a child of the Nature immersed into the
$\Psi$--ether, and it seems that he should possess the possibilities, yet not
discovered by science, of detecting the ether oscillations and of their
analysis. But it is very likely that the man is not perfect. Then, science is
faced with a honored responsibility to render assistance in the discovery of
new possibilities of the man up to his more full merging with the Universe
surrounding him.

Of course, in this way, the road opens towards occultism. But we should
temporarily reconcile ourselves to this, till scientific knowledge of the
$\Psi$--ether propeties establishes new data on the man's nature.

So, I assert that the $\Psi$--ether does exist! Further negation of it will
only hamper the development of physical, chemical, biological, and
philosophical knowledge.
\vspace*{5mm}

{\bf Conclusion}
\vspace*{5mm}

To summarize:

1. A historical review of the problem is given, from which it follows that
the $\Psi$--ether exists and its elimination from physical reality is
impossible, since without it, one cannot introduce the unity of measure and
number in the Universe, cannot physically substantiate the term
$\frac{l(l+1)}{r}$ in the radial part of the equation in the problem of a
hydrogen atom, and cannot physically explain the necessary requirement of
invariance of the equations of the Maxwell classical electrodynamics and
equations of the modern quantum field theory under the Lorentz transformations.
Just as the Galilei transformations (the Galilei principle of relativity) are
based on experimental data, so the Lorentz transformations are based on the
Michelson--Morly experimental data.

2. The model of
$\Psi$--ether is defined as the Bose--Einstein condensate of
neutrino--antineutrino pairs of the Cooper type; the physical grounds for the
model are given. The equation for the state of $\Psi$--ether as a continuous
medium is of the form
$$
\Box \Psi_i = 0,
$$
where $\Psi_i$ is a 4-dimensional vector.

The model provides propagation of transverse electromagnetic waves in the
$\Psi$--ether and waves of other types. It is not obligatory that the velocities of
propagation of different waves of the $\Psi$--ether would all be equal to the
light velocity.

A list of some properties of the
$\Psi$--ether is presented.

3. The limit of applicability of equations of classical and quantum
electrodynamics is formulated. These equations are valid as long as the
$\Psi$--ether is considered as a continuous medium where oscillations of the
$\Psi$--ether can propagate. However, when the wavelength of the $\Psi$--ether
becomes comparable with or smaller than sizes of the psyon, a molecule of the
$\Psi$--ether, or when in the region of long wavelengths, a wave-shaped
character of motions of the $\Psi$-ether is broken and transforms into a chaotic
motion of its large masses, the formation of electromagnetic waves ceased,
the man stops to register them with the optical instruments and radio
installations. The man does not "hear" the full voice of the Universe,
becomes "blind" and "deaf" in the Universe.

4. The contribution of the $\Psi$-ether can be detected, in particular, in
precision measurements of masses and lifetimes of a number of unstable
particles (hyperons, muons, pions, and other mesons) aimed at searching for
the "exact structure" of masses and lifetimes of those particles. These
experiments could shed new light upon the composition of particles from
protons, electrons, and neutrinos (and the corresponding antiparticles) into
which unstable particles ultimately decay. In this connection, the
confirmation of data obtained by the collaboration "KARMEN" is important.

5. The next step in cognition of secrets of the Universe and the secret of
life on the Earth is to study the properties of the $\Psi$--ether and its
constituents, psyons. This is evidently a task of physics of the XXIth
century.
\newpage

\begin{center}
{\bf References}
\end{center}
\begin{enumerate}

\item P.A.M.Dirac. "The Evolution of the Physicist's Picture of Nature".
Scientific American, May 1963, v.208, N 5, p.45--53.
\vspace*{-2ex}

\item J.Schwinger. A Magnetic Model of Matter. Science, 165 (No.3895),
757 (1969).
\vspace*{-2ex}

\item J.C.Maxwell. The scientific Papers. v.II. p.322.
Cambridge University Press, 1890.
\vspace*{-2ex}

\item J.C.Maxwell. The scientific Papers. v.II. p.763-765.
Cambridge University Press, 1890.
\vspace*{-2ex}

\item J.C.Maxwell. The scientific Papers. v.II. p.773-774.
Cambridge University Press, 1890.
\vspace*{-2ex}

\item Versuch einer Theorie der elektrischen und optischen Erscheinungen
in bewegten k\"orper. Leiden, 1895. H.A.Lorentz,  \S 89-92.
\vspace*{-2ex}

\item Henri Poincar\'e. Sur la dynamique de l'electron- C.R.Acad. Sci., 1905,
140, p.1504-1508; Oeuvres, t.IX.,p.489-493. Sur la dynamique de l'electron.-
Rendiconti Circolo mat. Palermo, 1906, 21, 129-176.
\vspace*{-2ex}

\item Henri Poincar\'e. La dynamique de l'electron- Revue g\'en. Sci. pures
et appl., 1908, 19, 386-402; Oeuvres, t.IX, p.551-586.
\vspace*{-2ex}

\item Henri Poincar\'e. L'hypoth\'ese des quanta. Revue scient.,
$4^{0}$ S\'er., 1912, 17, 225-232. Oeuvres, t.IX, p.654-668
\vspace*{-2ex}

\item "The principle of relativity" A collection of original memoirs on
the special and general theory of relativity by H.A.Lorentz, A.Einstein,
H.Minkowski and H.Weyl. Dover publications, inc.; p.37-38.
(Translated from "Zur  Electrodynamik bewegter K\"orper", Annalen der Physik,
17, 1905).
\vspace*{-2ex}

\item A.Enstein. Ather and Relativit\"atstheorie. Verlag von Julius Springer.
Berlin, 1920.
\vspace*{-2ex}

\item  W.Heitler. The Quantum theory of radiation. Third edition. Oxford.
At the Clarendon PRESS. 1954.  \S 1,6.
\vspace*{-2ex}

\item N.N.Bogolubov, D.V.Shirkov. Infroduction to the theory of Quantized
Fields. \\Interscience publishers, INC, New York, Interscience Puplishers Ltd.,
London, 1959,  \S6, p.59.
\vspace*{-2ex}

\item L.I.Schiff. Quantum Mechanics. Mc Graw-Hill Book Company, INC, New York,
Toronto, London, 1955,  \S14, p.74
\vspace*{-2ex}

\item Louis de Brogle "Problemes de propagations guidees des ondes
electromagnetiques". Paris, 1941, ch.III,  \S6.
\vspace*{-2ex}

\item Louis de Brogle. Henri Poincar\'e et les th\'eories de la physique.
Livre du Centenaire. Paris. Gauthier-Villars, 1956.
\vspace*{-2ex}

\item Louis de Brogle "Problems de propagations guidees  des ondes
electromagnetiques", Paris, 1941. ch.III,  \S6.
\vspace*{-2ex}

\item Phys. Lett., B348(1995), p.19.
\vspace*{-2ex}

\item S.N.Gninenko and N.V.Krasnikov. "Exotic muon decays and the KARMEN\\
anomaly", Phys. Lett. {\bf B434} (1998) 163; {\tt hep-ph/9804364}.
\end{enumerate}

\end{document}